\documentstyle[12pt,graphicx]{article}
\title{\bf Fundamental stellar parameters}
\author{M.~Wittkowski
\vspace{1cm}\\
\normalsize European Southern Observatory,\\ 
Karl-Schwarzschild-Strasse 2, 
85748 Garching bei M\"unchen, Germany,\\ 
e-mail: mwittkow@eso.org}
\date{\mbox{}}
\begin{document}
\maketitle
\pagestyle{empty}
%
%
\def\bull{\vrule height .9ex width .8ex depth -.1ex}
\makeatletter
\def\ps@plain{\let\@mkboth\gobbletwo
\def\@oddhead{}\def\@oddfoot{\hfil\tiny\bull\quad
``Science Case for Next Generation Optical/Infrared Interferometric Facilities 
(the post VLTI era)'';
37$^{\mbox{\rm th}}$ Li\`ege\ Int.\ Astroph.\ Coll., 2004\quad\bull}%
\def\@evenhead{}\let\@evenfoot\@oddfoot}
\makeatother
%
%
\def\beginrefer{\section*{References}%
\begin{quotation}\mbox{}\par}
\def\refer#1\par{{\setlength{\parindent}{-\leftmargin}\indent#1\par}}
\def\endrefer{\end{quotation}}
%
%
{\noindent\small{\bf Abstract:} 
I present a discussion of fundamental stellar parameters and their
observational determination in the context of interferometric
measurements with current and future optical/infrared interferometric
facilities. Stellar parameters and the importance
of their determination for stellar physics are discussed. 
One of the primary uses of interferometry in the field of stellar physics 
is the measurement of the intensity profile across the stellar disk, both
as a function of position angle and of wavelength. 
High-precision fundamental stellar parameters are also 
derived by characterizations of binary and multiple system using
interferometric observations. This topic is discussed 
in detail elsewhere in these proceedings. 
Comparison of observed spectrally dispersed
center-to-limb intensity variations with models of stellar atmospheres 
and stellar evolution may result in an improved understanding
of key phenomena in stellar astrophysics such as the precise evolutionary
effects on the main sequence, the evolution of metal-poor stars, 
stellar pulsation, mass-loss from high-mass main-sequence stars as well as 
from evolved stars, circumstellar environments, stellar magnetic activity, 
stellar rotation, and convection and turbulent mixing. Examples
of already achieved results with existing interferometric facilities
and anticipated improvements during the VLTI era are described.
A science case for a next-generation optical/infrared interferometric
facilities is presented, and the required specifications to achieve it
are given. Finally, important synergy effects with external facilities 
in order to reach a more complete picture are discussed.}
%
%
\section{Introduction}
High-spatial-resolution studies across the electro-magnetic 
spectrum directly revealing the intensity distribution, and not only 
the integrated flux density, of astrophysical objects are an essential 
tool to further our understanding in all parts of astrophysics. 

For a complete description of a star, a number of parameters are 
important, such as mass, luminosity, radius, age, pulsation period,
chemical composition, angular momentum, magnetic field, 
mass-loss rate, and the circumstellar environment. 
The observational determination of several of these parameters
requires a high spectral resolution. Furthermore, while some of
these parameters are integrated quantities, several of them
vary across the stellar disk and its environment, leading to the need
for spatially resolved observations. For example, the radius of 
a star is not a well-defined quantity since stars are gaseous
spheres and do not have a well-defined edge. Observable is in fact
the center-to-limb intensity profile across the stellar disk and 
its environment. For more detailed reviews on this aspects, see
for instance Scholz (1998, 2001, 2003).

Optical and infrared interferometry has already proven to be a
powerful tool for stellar astrophysics, in particular by providing 
fundamental stellar parameters such as CLVs (i.e., the stellar radius
in first order) and masses, which are compared to predictions by 
models of stellar evolution and stellar atmospheres. 

More detailed interferometric observations with current and future
facilities, coupled with a high spectral resolution, promise to
lead to a new level of accuracy of measured fundamental stellar parameters, 
and hence to a significantly improved understanding of 
stellar atmospheres and stellar evolution.
  
In this article, I discuss fundamental stellar parameters and the
importance of their measurement for stellar physics in the context
of measurements with current and future interferometric facilities.
In Section 2, I discuss fundamental parameters and their relation
to directly observable parameters. In Sect. 3, I remind of a few examples of 
interferometric observations already achieved in this field.
A discussion of the anticipated progress of interferometric
measurements making use of the facilities in the VLTI era is
provided in Sect. 4. In Sect. 5, I present a science case
for a next-generation optical/infrared interferometric facility
that can not already be reached with the facilities of 
the VLTI era. Specifications for such a facility are derived.
Finally, since optical/infrared interferometry alone is not able to provide
the complete set of parameters that fully describe a star and its
environment, important synergies with external facilities in order to 
reach a more complete picture are discussed in Sect. 6.

Interferometric observations of binaries and multiple systems
provide as well fundamental stellar parameters, most importantly
precise stellar masses. This aspect is in detail discussed by
F. Verbunt (these proceedings), and therefore mostly left out
in this article.

This article is not meant as a review of the field. In particular,
the mentioned theoretical models, and the presented observations
are just meant as examples. Many more interesting results have been
achieved.
For recent reviews of the topics of interferometric techniques,
including their application to fundamental stellar parameters, see 
for instance, Quirrenbach (2001) and Monnier (2003). For a review
on ``Accurate masses and radii of normal stars'', see Andersen (1991). 
Also, various lecture notes  from Michelson Summer Schools on 
interferometry include presentations of fundamental stellar parameters 
in the context of optical/infrared interferometry, for example those by 
John Davis, Michael Scholz, Mel Dyck, Christian Hummel (2000); 
John Davis, Jason Aufdenberg (2003).
\section{Fundamental Stellar Parameters}
\subsection{Fundamental parameters in stellar physics}
Models of stellar evolution and stellar atmospheres are based
on a set of fundamental parameters that describe the (model) star.
\paragraph{First-order fundamental parameters}
The mass $M$, the luminosity $L$, and the radius $R$ are fundamental
parameters of any star. In addition, the chemical 
composition is needed to describe a star. These quantities
evolve as a function of time, where the zero-point is usually defined
when the star appears first on the main sequence, i.e.
the so so-called zero age main sequence (ZAMS). For comparisons of 
observations to theoretical models, the current age of the observed 
star with respect to the time grid of the model is thus an additional 
a priori unknown parameter. 

The effective temperature 
$T_{\rm eff}=L/4\pi R^2 \sigma$,
the surface gravity $g=GM/R^2$, and the mean density
$\overline{\rho}= 3M/4\pi R^3$ are dependent parameters of the set
\{$M$, $L$, $R$\}, that are sometimes used to replace one or more of the 
set of parameters \{$M$, $L$, or $R$\} for reasons of convenience and 
in order to use a parameter set that may be closer to directly 
observed quantities.

Theoretical evolutionary calculations are often shown as 
mass-dependent tracks or isochrones in a (theoretical) 
Hertzsprung-Russel diagram ($\log L - \log T_{\rm eff}$ diagram).
Mass-luminosity- ($L-M$) and mass-radius- ($M-R$) relations are also
used for comparisons with observations.

Recent grids of stellar evolutionary calculations were made available,
for instance, by Girardi (2000) or Yi et al. (2003). These
works include figures of mass tracks and isochrones in 
theoretical Hertzsprung-Russel diagrams. Theoretical
mass tracks are for instance used to estimate the mass of a star
based on measured effective temperature and luminosity. Isochrones
are for instance used to estimate the age of a globular cluster.
Girardi (2000) present as well a discussion on the precise effects
on the main sequence as a function of stellar metallicity.
Such small effects are so far not well constrained by observations
due to their insufficient precision. This is for instance a field
where a next-generation optical/infrared interferometric facility
could have a significant impact on evolutionary calculations. 

\paragraph{Additional fundamental parameters}
The first-order set of fundamental parameters discussed above
may not be sufficient for more detailed studies of stellar
physics, especially if stars in certain evolutionary stages
are considered. The following parameters can be
of fundamental importance for a complete description of a star 
as well.
These parameters include the (initial) angular momentum $I$, 
the magnetic field $B$, the mass loss rate $\dot{M}$, the pulsation 
period $P$, and also the characteristics of the circumstellar environment.

Examples of the non-negligible effects of rotation and 
magnetic fields on evolutionary tracks, for instance in a 
$\log L - \log T_{\rm eff}$ diagram, were recently presented
by Maeder \& Meynet (2003, 2004). Effects of the time-dependent
mass loss rate on the evolutionary tracks, together with comparison
to observations, are shown in Girardi et al. (2000).

\paragraph{Purpose of measured fundamental stellar parameters}
A next generation optical/infrared interferometric facility can provide 
measurements of fundamental parameters with a precision beyond the
capabilities of current facilities and of the ``VLTI era''. Also,
a more complete set of measured fundamental parameters can then be reached.  
There are several purposes for which an improved knowledge of
fundamental stellar parameters is required in order to
improve our knowledge of stellar physics. 

Complete tests of stellar evolution theory and as well as of stellar
atmosphere theory require high-precision measured fundamental
parameters for stars in all evolutionary stages across the HR-diagram.
In particular for small stars with low apparent magnitude, there
are so far not many observational constraints. For all stars, the
accuracy of obtained fundamental parameters depend also on observational
constraints of additional effects on the stellar surface. For instance
close circumstellar matter or surface spots may easily bias measurements
of stellar radii if those effects are not simultaneously probed by the
observations.

Constraints of the precise effects on the main sequence, as mentioned
above need very high-precision effective temperatures, luminosities,
and chemical compositions. Furthermore, main sequence stars have
usually small angular diameters that require long baselines for precise
measurements of their radii.

Different available material tables such as opacity tables, 
nuclear reaction rates, etc. can be tested only with higher 
accuracy measurements than currently available.

The physical description of key phenomena such as convection,
mass loss, turbulent mixing, rotation, magnetic activity,
can be significantly improved with more detailed interferometric 
measurements than currently available .
\subsection{Relation of fundamental parameters to observables}
Many of the (theoretical) parameters mentioned above, which are 
important quantities for theoretical calculations, can not easily be 
measured directly as such, but are deferred from measurements
of other observational quantities.

Moreover, some of the (theoretically defined) parameters mentioned 
above are not well-defined parameters when additional effects
for observed stars are considered. The radius $R$, for instance, 
is not well defined for observations. Stars are gaseous spheres, 
and do not have a well-defined ``edge''. Observable is the 
center-to-limb intensity variation (CLV) across the stellar disk 
(and its circumstellar environment), which depends on the 
star's atmospheric structure (for detailed reviews on this aspect, 
see e.g., Scholz 1999, 2001, 2003). The CLV is a measure of the
vertical temperature profile of the stellar atmosphere. It may be
superimposed by horizontal surface inhomogeneities.
Also, quantities such as magnetic fields and chemical composition
may not be constant, but may vary across the surface of the star.
It is, thus, much more accurate to compare observed and model-predicted
CLVs rather than just radii.

The mass $M$ and luminosity $L$ are well-defined integrated
physical parameters. Since these are integrated 
quantities, their measurement does not necessarily require a 
very high angular resolution. A spatial resolution below  the order 
of the angular size of the stellar disk is usually sufficient.
\paragraph{Mass $M$} Binary stars are the main source of
measurements of high-precision masses. One can distinguish between 
double-lined astrometric-spectroscopic binaries 
(measurement of the radial velocities of both components
and of the relative orbit), absolute astrometry (measurement of the
absolute orbits for both stars), and single-lined spectroscopic 
binaries (measurement of the radial velocity of one component)
together with a measurement of the relative orbit and the distance.
Interferometry can surely improve high-precision mass estimates
by measurements of high-accuracy astrometric orbits. 
The topic of binary and multiple systems is in detail 
presented by F. Verbunt (these proceedings), it will not be 
covered further in this article.

In addition, the surface gravity and effective temperature
can indirectly be measured by means of spectroscopic observations.
Coupling interferometric and spectroscopic techniques, these
quantities can be derived with higher precision by comparing observed
and model-predicted CLVs in the continuum and certain (narrow) lines.  
This approach makes use of the effect of stellar mass and (surface) 
temperature on the atmospheric structure. Together with a measurement 
of the stellar radius (from measured angular diameter and distance), the 
surface gravity gives the stellar mass. This mass estimate is, thus, 
not a direct measurement. However, this is a very interesting check of
the consistency of observations with stellar evolution theory and
stellar atmosphere theory.
\paragraph{Luminosity $L$} The luminosity can be derived from 
observations of the apparent bolometric flux and a distance estimate, 
such as the Hipparcos parallax. Accuracies of luminosities are
currently often limited by the accuracies of the distances.

For improved observations of this parameter, higher accuracy
distances (parallaxes) are required, as well as a high-precision
bolometric photometry. For variable stars, the bolometric flux
has to be obtained at the same time as other parameters such as
the radius. This is, for instance, an important aspect where 
carefully planned synergies with external facilities may be highly
valuable.
\paragraph{Radius $R$} The radius is often measured by means of
optical interferometry. As mentioned above, the radius is not a 
well defined quantity. The direct observable is the CLV, which may
be described by a parametrization. One resolution element across 
the stellar disk allows us usually to constrain only one parameter. 
In other words, one resolution element gives usually only the 
angular diameter for a a-priori adopted CLV model. 
Two resolution elements, usually including measurements of the visibility 
function in the second lobe, result in an estimate of the angular 
diameter and one additional parameter that described the shape
of the CLV. Different radius definitions as for instance the 
Rosseland mean diameter, the continuum diameter, or an equivalent
UD diameter do not necessarily coincide. Often, uniform disk angular 
diameters in a particular bandpass are measured and transformed into 
Rosseland mean diameters or continuum diameters using 
atmosphere models.
For compact atmospheres, for instance for regular main sequence 
stars, different radius definitions may agree within the precision
of current observations. 
For extended atmospheres, for instance for giant stars, these
radii differ already within the precision of current observations
(see, for instance, Wittkowski et al. (2001, 2003); or the reviews
by Scholz (1999, 2001, 2003). 
CLV measurements at various narrow wavelength intervals will give the 
best characterization of the stellar radius by comparison with
model-predicted CLVs. With improved
observational accuracy, this approach may even result in a 
direct determination of the vertical and horizontal temperature 
profile. The observations by Perrin et al. (2004) 
demonstrate the importance of radius observations in narrow
wavelength bands.
\paragraph{Chemical composition}
At the given evolutionary stage of an observed star, the chemical
composition on its surface is observable, for instance by using
high-spectral-resolution observations. However, if one wants
to measure abundance surface inhomogeneities, which may for instance
be caused by magnetic fields, the chemical composition needs to be 
determined for several resolution elements across the stellar disk.  
In this respect, one future objective might be to 
discriminate stellar tracks with different initial chemical
composition, and another to probe abundance inhomogeneities
across the stellar surface.
\paragraph{Age} The age of a star is extremely difficult to be
measured directly. It is usually obtained by comparing 
stellar evolutionary tracks with observational estimates
of mass, luminosity, and radius. A test of the time dependence of 
evolutionary tracks would require an independent age determination. 
One approach to reach this is to identify stars with identical
age, such as binary and multiple systems. An absolute age determination
of a globular cluster was recently obtained by observations of
Berillium spectral lines by Pasquini et al. (2004). 
\paragraph{Angular momentum} For rotating stars, the 
(initial) angular momentum is an important fundamental parameter for the 
stellar model. Observable is the surface rotational velocity at the 
evolutionary stage of the observed star, 
by means of spectroscopic observations. Again, spatially resolved
observations are needed to constrain the rotational parameters 
in more detail.
\paragraph{Magnetic fields} The magnetic field at the surface can 
be measured by observations of line splittings. The field strength
may vary across the disk, again requiring a coupling of high
spatial and high spectral resolution. 
\paragraph{Mass-loss rate} The mass-loss rate can for instance be 
derived from outflow velocities.
\paragraph{Pulsation period} The pulsation period is an 
important parameter for variable stars, such as Cepheid or Mira
variables. 
The variability of the luminosity, derived from monitoring 
of the bolometric flux, and the variability of the radius may 
differ in phase and even period. 
Different radius definitions, such as continuum and Rosseland-mean
radius can as well lead to different observed phases and periods.
See, for instance, the recent model studies by Ireland et al. (2004a,b). 
Luminosity-period relationships (of Cepheids) are of essential 
importance as distance indicator.
\subsection{Required Precision}
Referring to Andersen (1991), the precision needed for constraining
theoretical models varies for the different
parameters. For the mass and the radius an accuracy of 1\% to 2\%
is currently required, whereas for the luminosity a precision
of 5\%-10\% is sufficient. Higher accuracies are currently not needed
since other effects would then limit the computations, such as 
effects of the metallicity. However, as our understanding of these
limiting factors increases, a more sophisticated analysis of these
stellar parameters becomes necessary again. Thus, it seems worth
to carry out specific model investigations in parallel to the planning
of a next generation interferometric facility in order to investigate
the precision of the various parameters that will then be needed.
\subsection{Specifications I}
Resulting from this Section, a number of specifications for a
next generation interferometric facility can already be summarized.
\begin{itemize}
\item Radius measurements, i.e. measurements of the CLV and comparison
with model-predicted CLVs with 
sufficient accuracy require at least 2--3 resolution elements 
across the stellar disk.
\item CLV observations in narrow spectral bands (spectral resolution
larger than 100\,000) are needed to constrain spatially resolved
metallicity and magnetic fields.
\item In order to distinguish the CLV caused by the vertical
temperature profile and additional horizontal surface inhomogeneities,
an even higher number of resolution elements (5--10) may be necessary.
The aspect of imaging of stellar surfaces is discussed in detail
by von der L\"uhe (these proceedings).
Measurements of phases are essential to constrain
asymmetric intensity profiles.
\item For a complete understanding of stellar physics, high-precision
measurements of various stars across the Hertzsprung-Russell diagram
are needed, including stars with small angular diameters and low
luminosity. The next generation interferometer should provide baselines
and collecting areas that allow us to perform such observations.
\item In addition to interferometric measurements, additional
observational information is needed, such as high-precision
distances and bolometric fluxes. Carefully planned synergies with 
other facilities may be very interesting in order to reach this.
\end{itemize}
\section{Examples of interferometric results in stellar physics}
A few examples of interferometric observations in the
field of stellar physics are discussed below, in order to illustrate
the capabilities of current interferometric facilities.
Note that this discussion is not intended to have the character 
of a review. Many more exciting results have been obtained, see for 
instance the reviews by Quirrenbach (2001) and Monnier (2003), as well as 
the regularly updated newsletter OLBIN (olbin.jpl.nasa.gov), edited
by Peter Lawson.
\paragraph{Fundamental parameters of very low mass stars}
Low mass stars are characterized by a low effective temperatures and high
surface gravities. Because of the rather small angular sizes, these
objects are difficult for interferometric observations. However, first
measurements of diameters of M dwarfs succeeded already
(Lane et al. 2001, Segransan et al. 2003). These observations
provide an empirical mass-radius relation, which is compared to
theoretical stellar evolution calculation. These observations provide
also observational constraints for atmosphere models.
\paragraph{Effects of stellar rotation}
Interferometric measurements can also derive asymmetric shapes of
stellar surfaces if baselines of different orientation are used.
Van Belle (2001) presented recently the  detection
of an oblate photosphere of a fast rotator, the main sequence star
Altair. Domiciano de Souza et al. (2003) reported on VLTI measurements
of the asymmetric shape of the rotating Be star Achernar, and found 
that it is much flatter than theoretically expected.
\paragraph{Intensity profiles and limb-darkening}
Optical/infrared interferometry has proved its capability to reach beyond the
measurement of diameters, and to measure additional surface structure
parameters. Through the direct measurement of stellar limb-darkening,
interferometry tests the wavelength-dependent intensity profile across
the stellar disk. However, the required direct measurements of stellar 
intensity profiles are among the most challenging programs in current 
optical interferometry. Since more than one resolution element across 
the stellar disk is needed to determine surface structure parameters 
beyond diameters, the long baselines needed to obtain this resolution 
also produce very low visibility amplitudes corresponding to vanishing 
fringe contrasts. Consequently, direct interferometric limb-darkening 
observations of stars with compact atmospheres, i.e. visibility 
measurements in the 2nd lobe, have so far been limited to a small number 
of stars (including Hanbury Brown et al. 1974;
Di Benedetto \& Foy 1986;
Quirrenbach et al. 1996;
Burns et al. 1997;
Hajian et al. 1998;
Wittkowski et al. 2001, 2004).
For stars with extended atmospheres, described by for instance
Gaussian-type or two-component-type CLVs, measurements of high to medium
spatial frequencies of the 1st lobe of the visibility function
may lead to CLV constraints as well
(see, e.g. Haniff et al. 1995; Perrin et al. 1999;
Woodruff et al. 2004; Fedele et al. 2005). 

\begin{figure}
\centering
\resizebox{0.32\hsize}{!}{\includegraphics{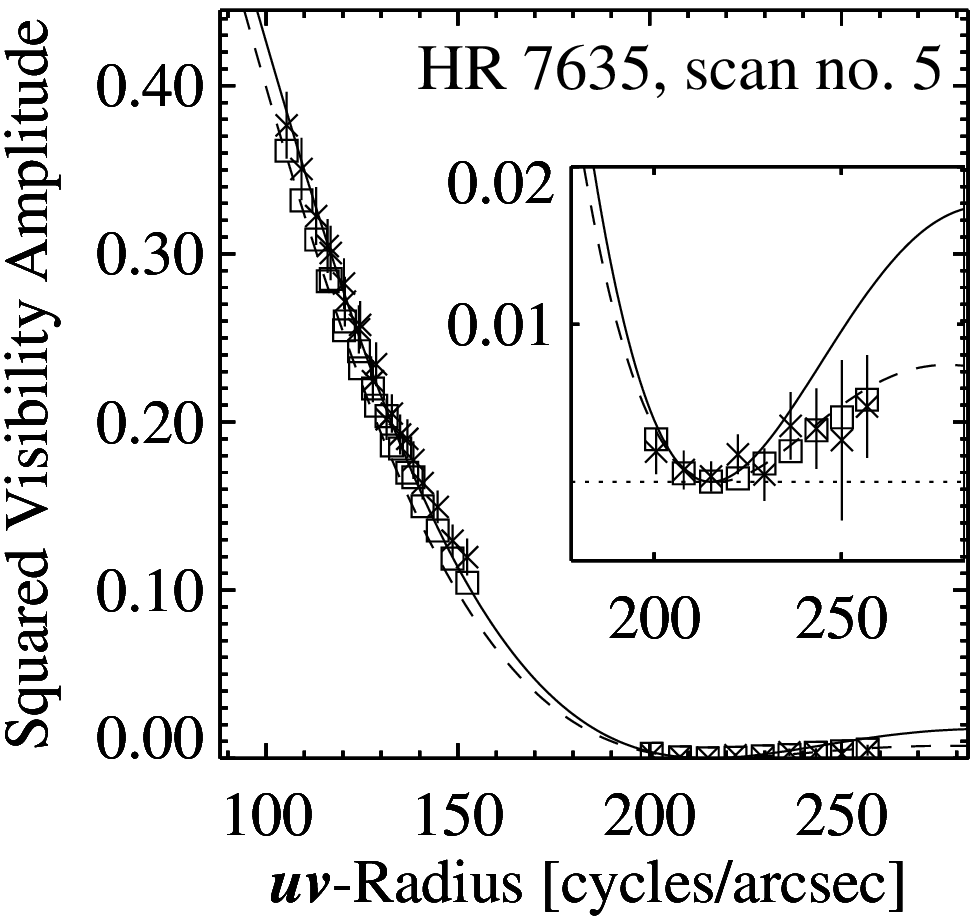}}
\resizebox{0.32\hsize}{!}{\includegraphics{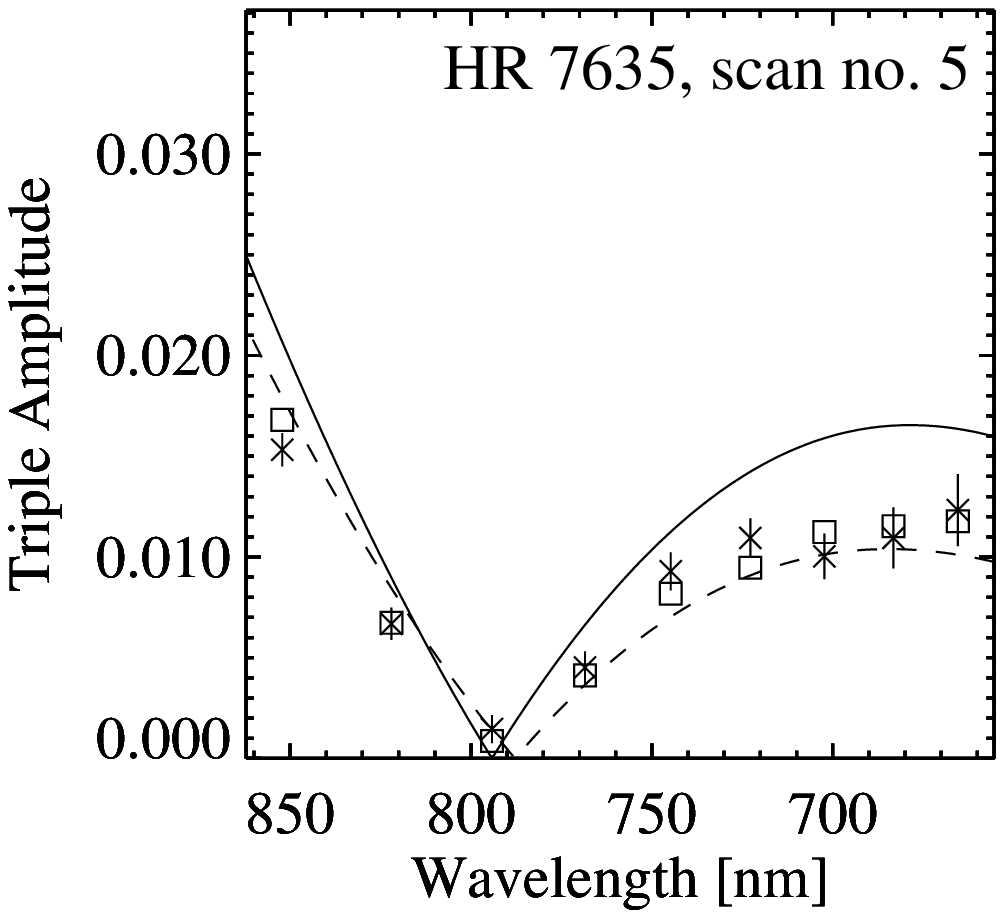}}
\resizebox{0.32\hsize}{!}{\includegraphics{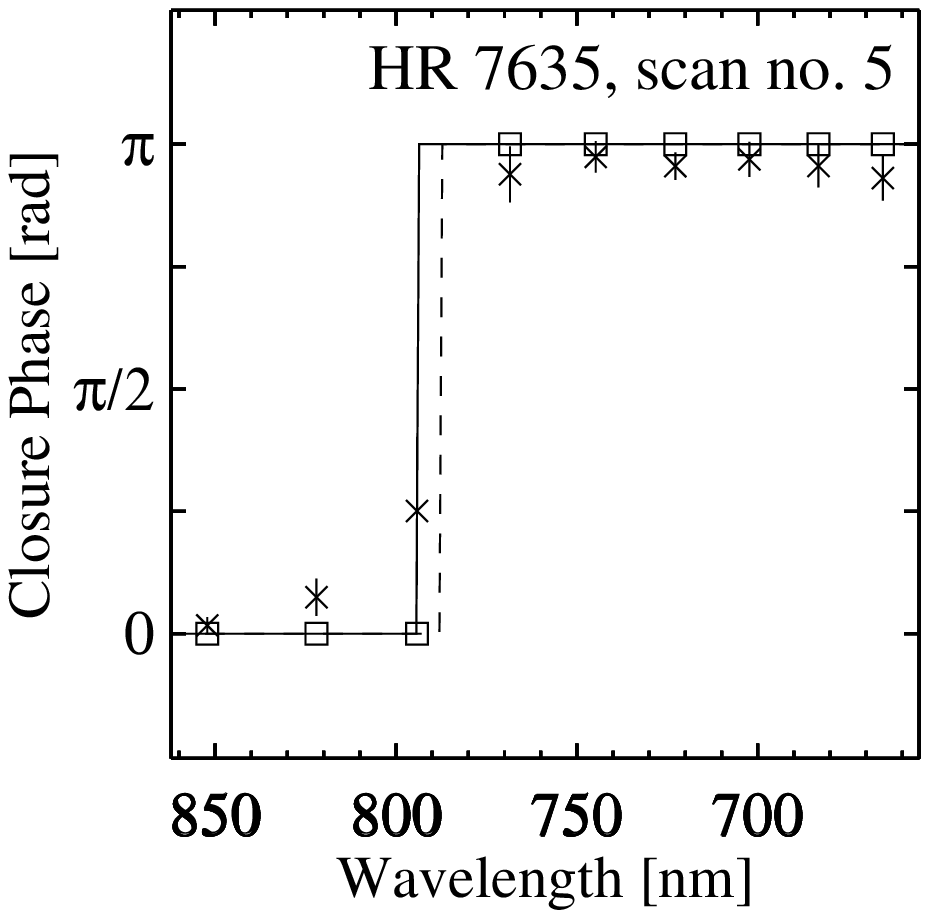}}
\caption{NPOI limb-darkening observations of the cool giant $\gamma$\,Sge 
(Wittkowski et al. 2001).}
\label{fig:gamsge}
\end{figure}

Fig.~\ref{fig:gamsge} shows an example of limb-darkening observations
of the cool giant $\gamma$\,Sge obtained with the NPOI (Wittkowski et al. 2001). 
These observations include squared visibility amplitudes, triple products, 
and closure phases for 10 spectral channels. They succeeded not only in 
directly detecting the limb-darkening effect, but also in 
constraining {\tt ATLAS\,9} model atmosphere parameters.

\begin{figure}
\centering
\resizebox{0.45\hsize}{!}{\includegraphics{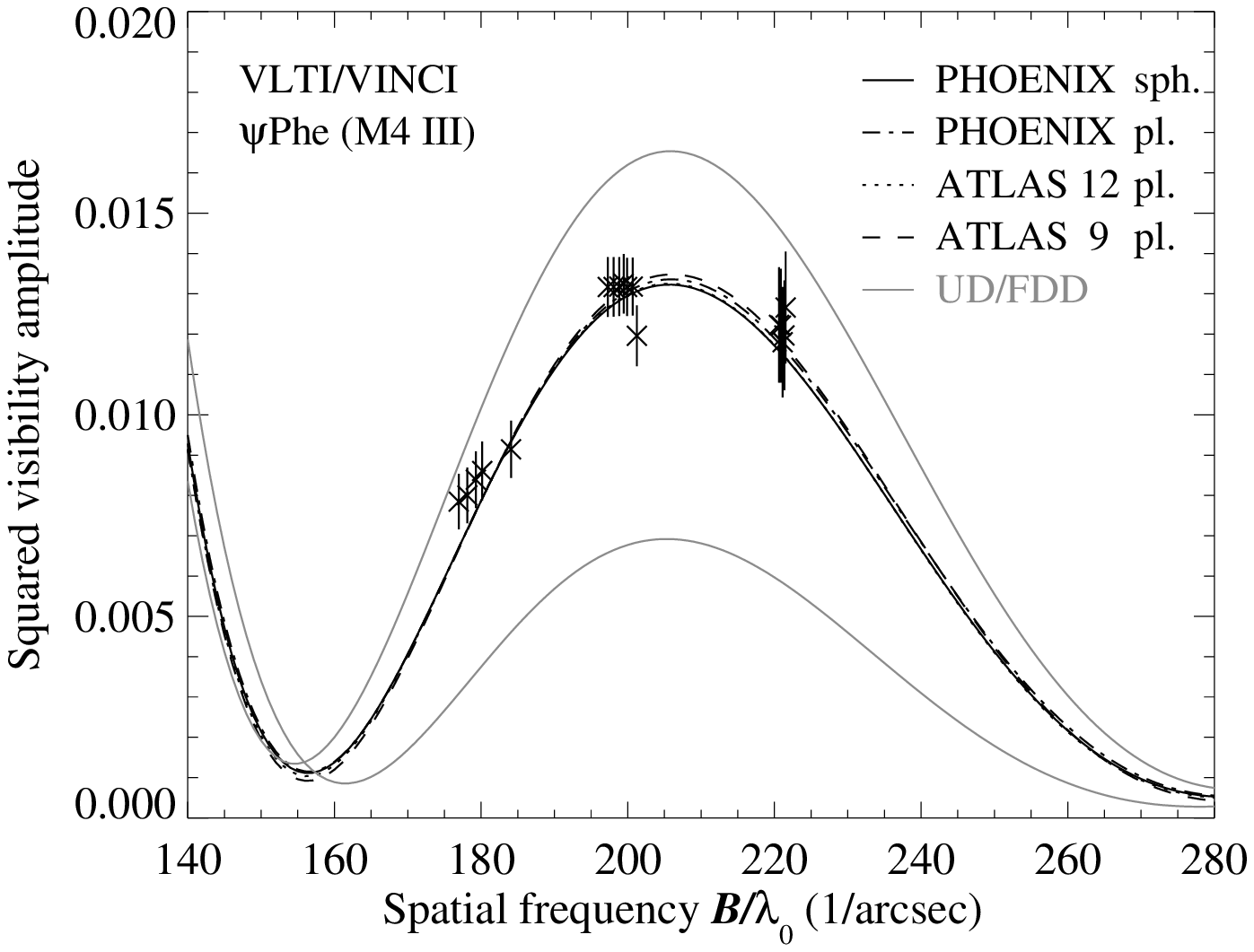}}
\resizebox{0.53\hsize}{!}{\includegraphics[angle=90]{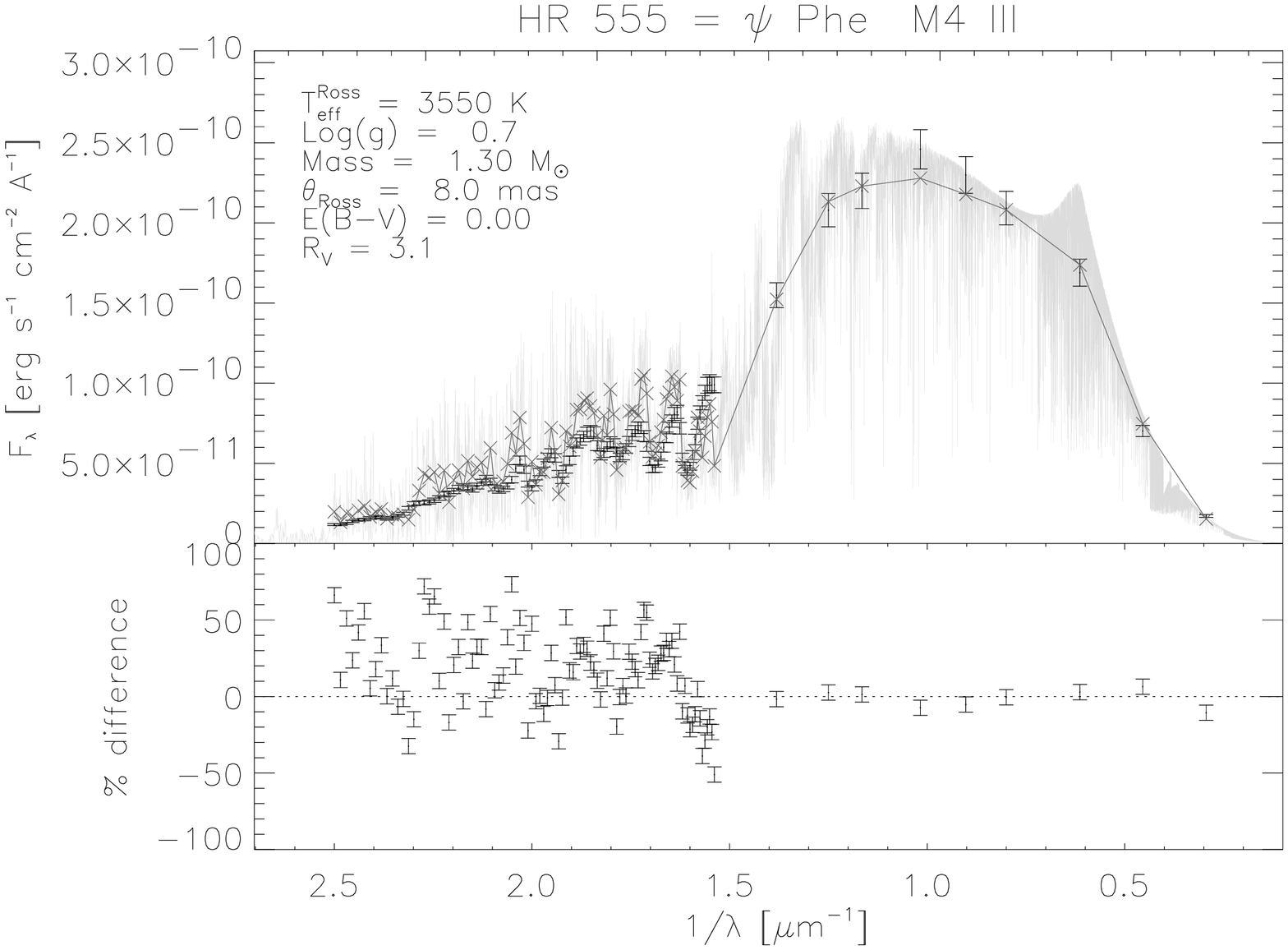}}
\caption{VLTI limb-darkening observations of the M4 giant $\psi$\,Phe 
(Wittkowski et al. 2004).}
\label{fig:psipheobs}
\end{figure}

\begin{figure}
\centering
\resizebox{0.7\hsize}{!}{\includegraphics{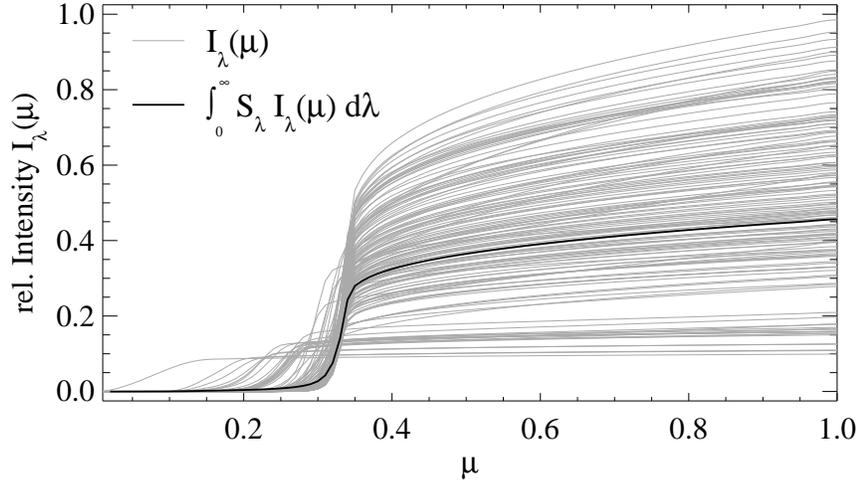}}
\caption{Monochromatic and filter-averaged model CLV for the M4 
giant $\psi$\,Phe (Wittkowski et al. 2004).}
\label{fig:psipheprof}
\end{figure}

Fig.~\ref{fig:psipheobs} shows an example of a limb-darkening observation
of the cool giant $\psi$\,Phe, obtained with the VLTI 
(Wittkowski et al. 2004). These observations were compared to {\tt PHOENIX}
and {\tt ATLAS} model atmospheres, which were also created by comparison
to stellar spectra (right panel of the Fig.). Fig.~\ref{fig:psipheprof}
shows for illustration the monochromatic and filter-averaged CLV of the 
well fitting {\tt PHOENIX} model atmosphere. These limb-darkening
observations also result in a very precise and accurate radius estimate 
because of the precise description of the CLV.
\begin{figure}
\centering
\resizebox{0.48\hsize}{!}{\includegraphics{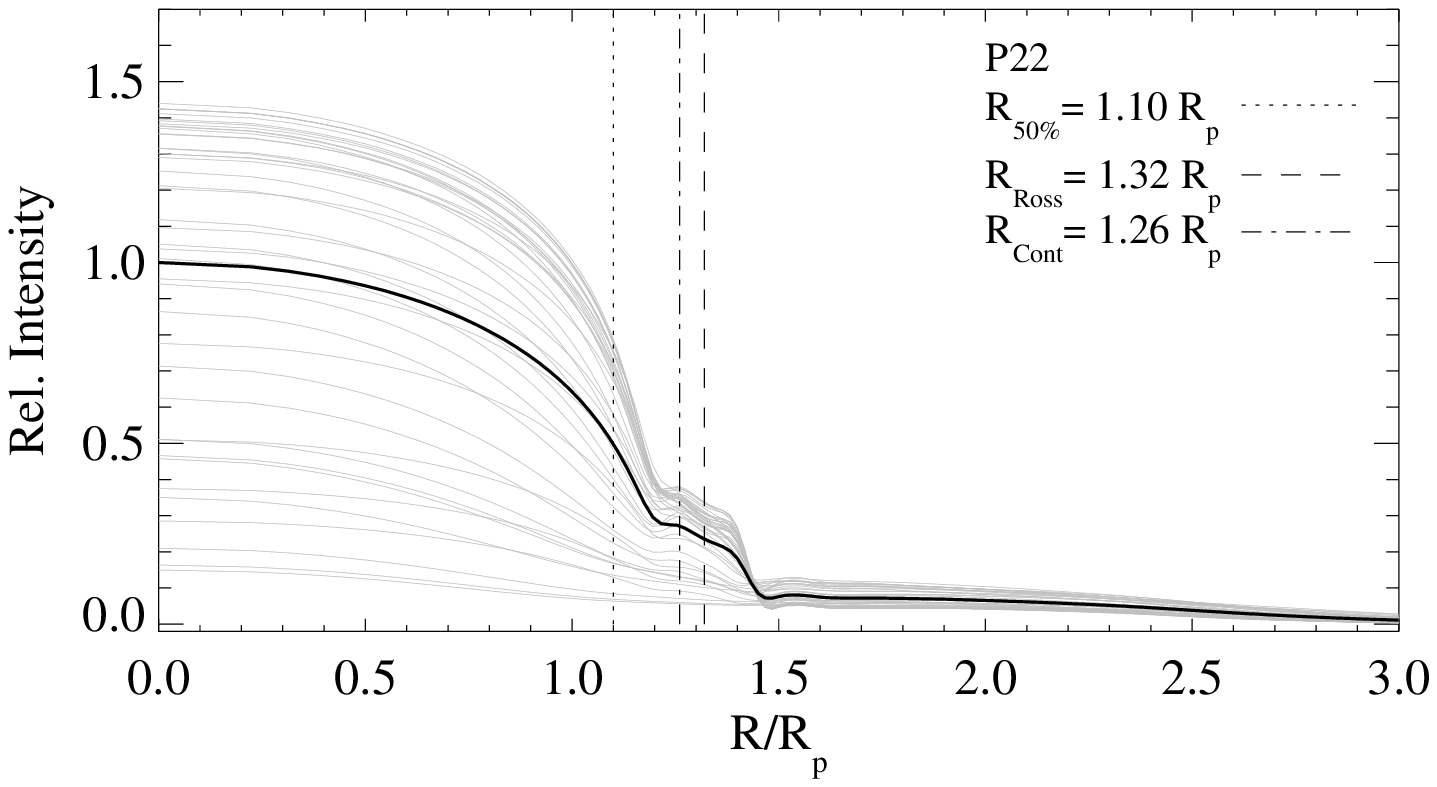}}
\resizebox{0.48\hsize}{!}{\includegraphics{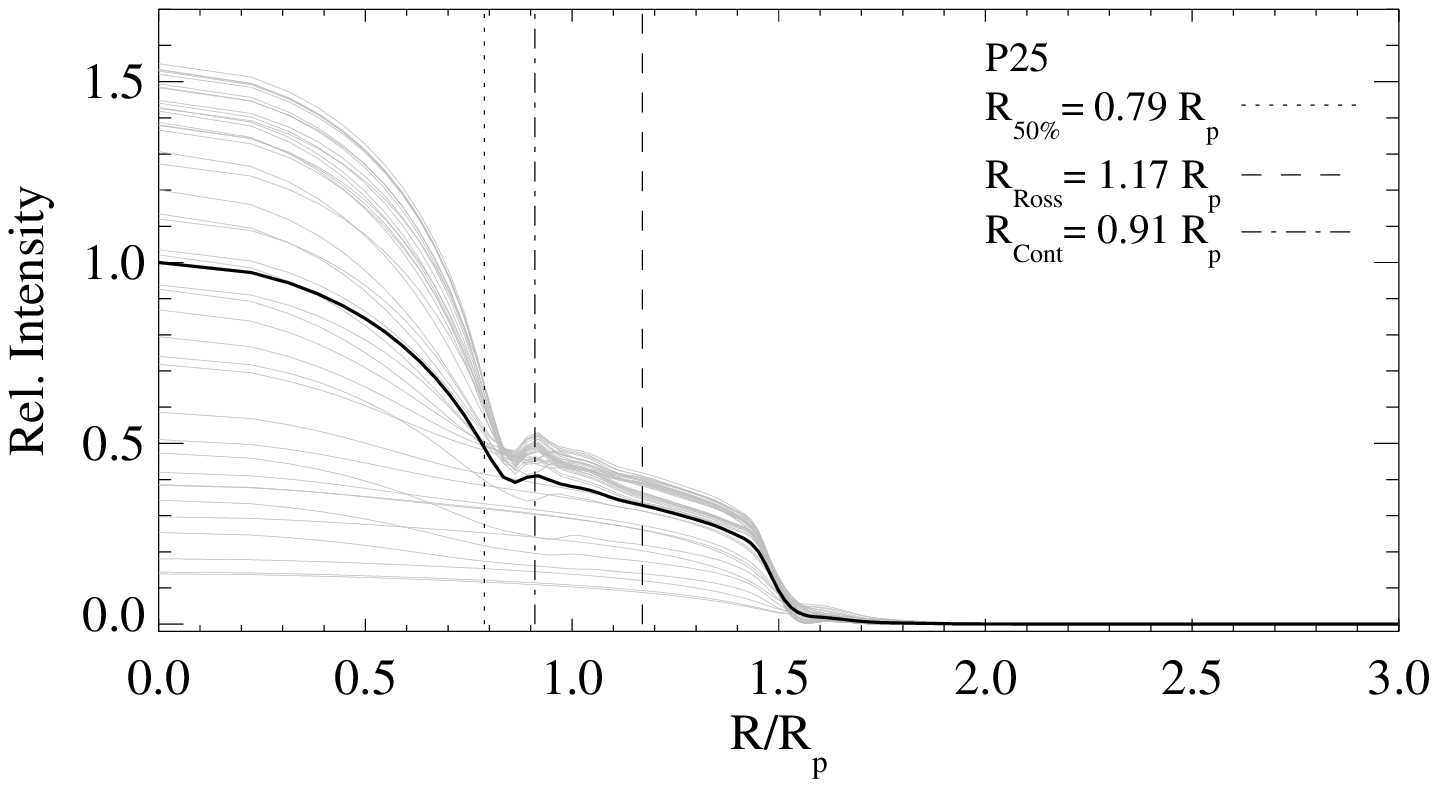}}

\resizebox{0.48\hsize}{!}{\includegraphics{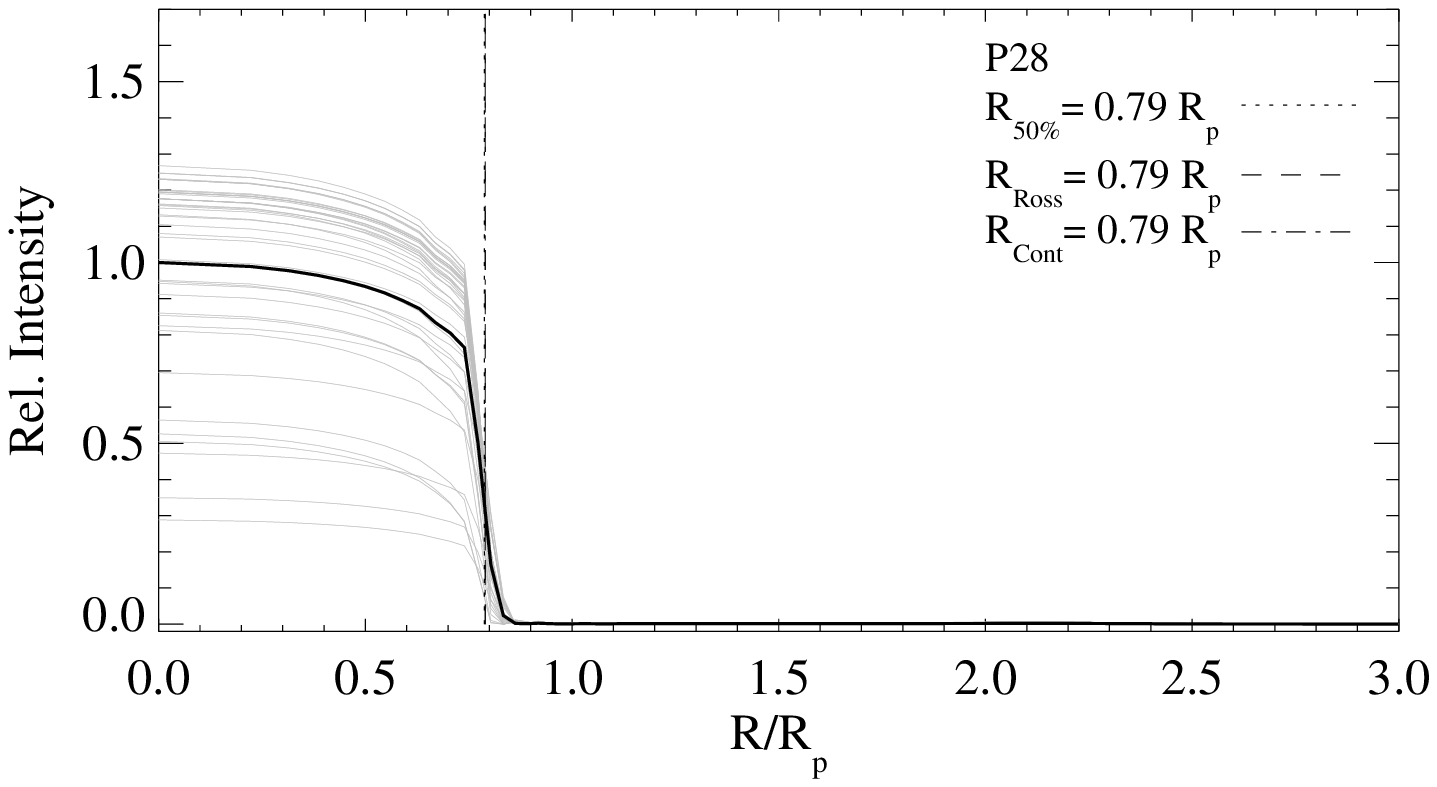}}
\resizebox{0.48\hsize}{!}{\includegraphics{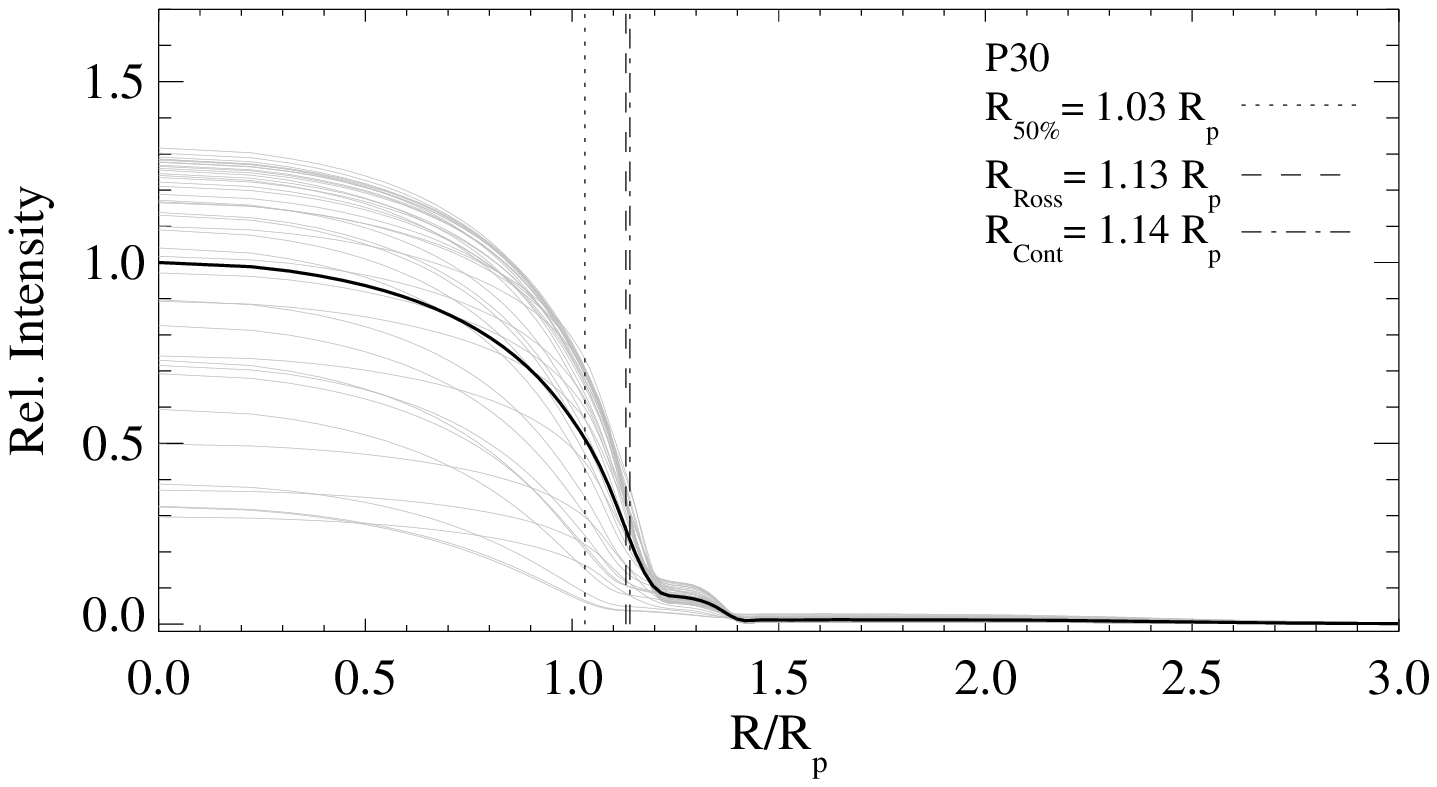}}
\caption{Monochromatic and filter-averaged model CLVs of Mira 
stars as a function of stellar phase, predicted  by dynamic model 
atmospheres (P series) described in Hofmann et al. (1998), 
Tej et al. (2001), Ireland et al. (2004a,b).
The model files used are from Scholz \& Wood (2004, private 
communication).}
\label{fig:profiles}
\end{figure}
\begin{figure}
\centering
\resizebox{0.48\hsize}{!}{\includegraphics{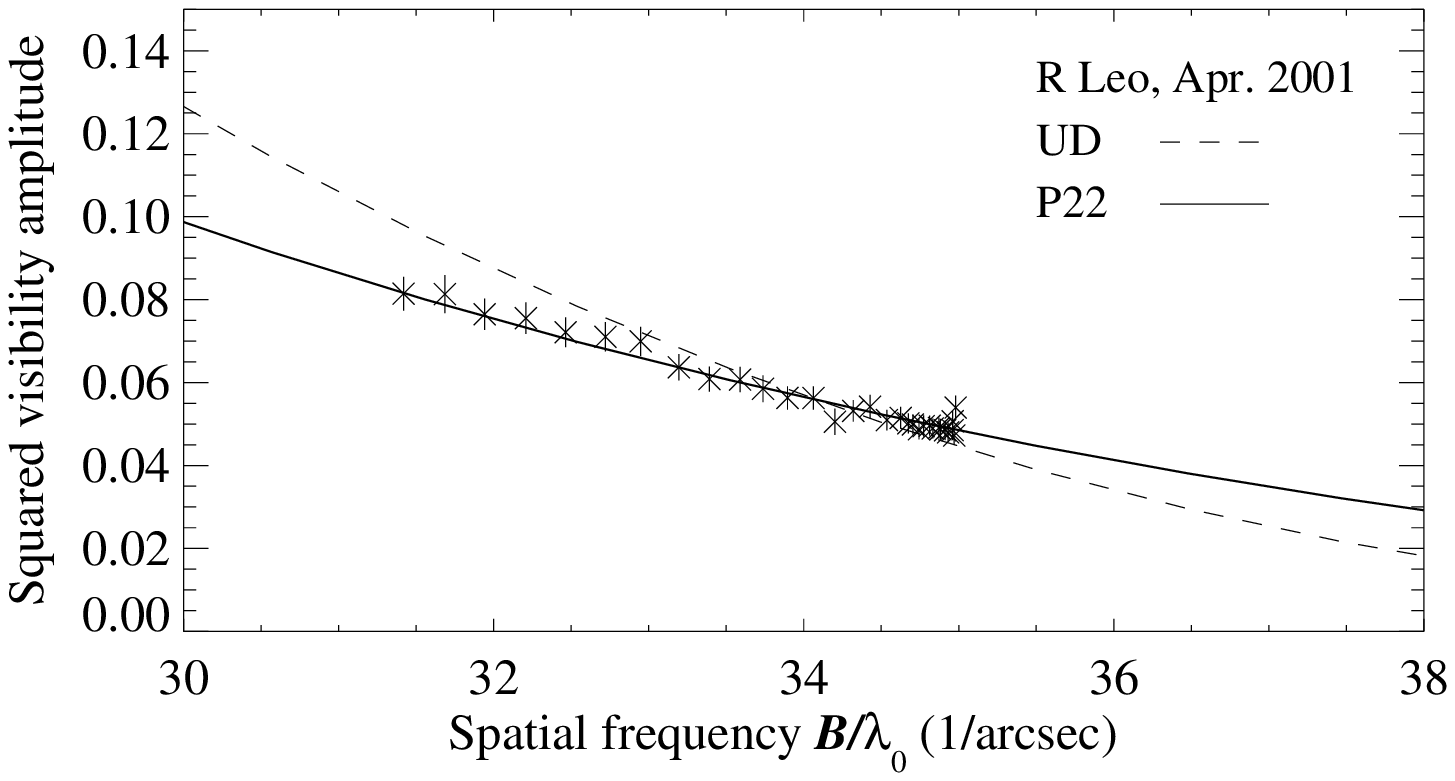}}
\resizebox{0.48\hsize}{!}{\includegraphics{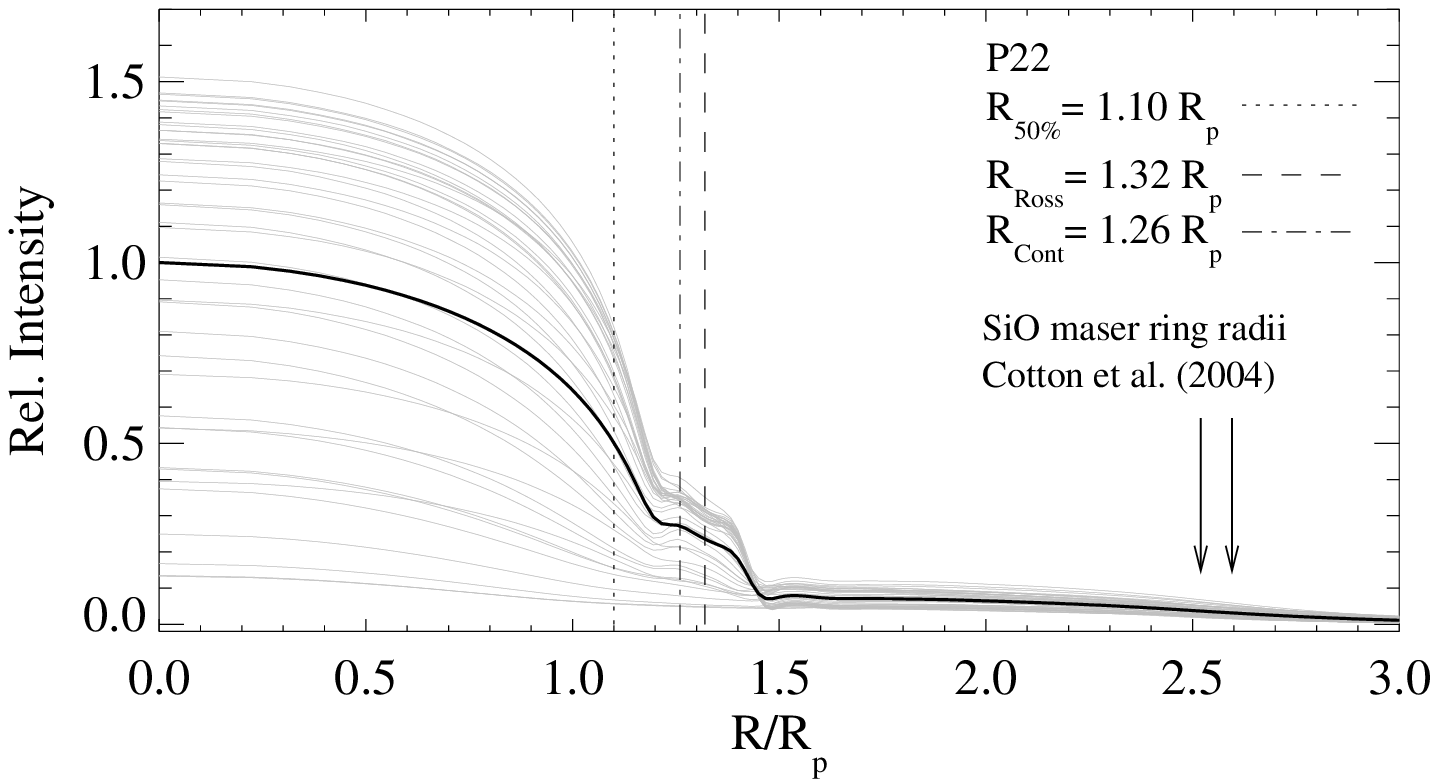}}
\caption{VLTI/VINCI observations of the intensity profile of the Mira
variable R Leo and comparison to dynamic model atmosphere predictions
(P series) by Hofmann et al. (1998), Tej et al. (2001), 
Ireland et al. (2004a,b).
The left panel shows a comparison of the visibility values to 
a well-fitting model (P22), and the right panel shows the CLV 
prediction by this model atmosphere. Also shown are the mean 
positions of the SiO maser shells observed by Cotton et al. 
(2004). From Fedele et al. (2005).}
\label{fig:rleo}
\end{figure}

For cool pulsating Mira stars, the CLVs are expected to be more complex,
which can be described as maybe Gaussian-shaped and more-component CLVs, 
and to be highly variable as a function of stellar variability phase. 
Fig.~\ref{fig:profiles} shows model predictions for Mira star CLVs
at four different variability phases from the hydrodynamic model atmospheres
by Hofmann et al. (1998), Tej et al. (2001), Ireland et al. (2004a,b).
The model files are from Scholz \& Wood (2004, private communication).
These complex shapes of the CLV make it even more difficult to define an
appropriate stellar radius. Fig.~\ref{fig:profiles} shows as well the
location of different radius definitions, the Rosseland mean radius, the
continuum radius, and the radius at which the filter-averaged intensity
drops by 50\%. For complex CLVs at certain variability phases these 
definitions can result in differences of up to about 20\%.  
Fig.~\ref{fig:rleo} shows VLTI observations of the Mira star R Leo 
compared to the dynamic model atmospheres described above
(Fedele et al. 2005). The direct 
comparison of measured intensity
{\it profiles} with the model CLVs allow us to clearly relate certain points
of the model CLVs to observable angular scales. With other words, a 
well-defined measurement of angular stellar sizes, and hence tests
of model atmospheres and model evolutionary calculations, becomes possible 
when observed CLVs are compared to model-predicted CLVs, despite the general
uncertainty of radius definitions. This, however, requires usually several
resolution elements across the stellar disk. 
\section{The VLTI era}
The completed VLTI will provide the astronomical community with
additional observational capabilities which are not yet available. 
Furthermore, other interferometric facilities will as well reach an 
increased level of observational possibilities during the same period
of time. With increased observational capabilities, also the 
state-of-the-art of astrophysical interpretation will improve during 
the next years.  The expected instrumental level during the ``VLTI era''
includes
\begin{itemize}
\item Baselines between 8\,m to 200\,m for the VLTI, and maybe up to the 
order of 1\,km with other facilities during this era.
\item A maximum collecting area corresponding to 8--10\,m telescopes.
\item A wavelength coverage from about 1$\mu$m to 10--20$\mu$m 
for the VLTI, and optical wavelengths with other facilities during this era.
\item Three-way beam combination with first-generation instrumentation,
up to 6--8-way beam combination with second generation (VLTI).
\item A dual feed phase referencing system (PRIMA at VLTI).
\end{itemize}
\section{The science case}
A strong general science case to justify
the construction of a next generation interferometric facility is 
presented below. This science case is based on the discussion in Sect. 2.
It is challenging, and clearly reaches beyond the observational 
capabilities of the VLTI era as outlined in Sect. 3. 
However, it appears realistic for the next generation facility. 
The presentation of the general approach is followed by
more specialized limiting cases that better allows us to 
specify the characteristics needed for the next generation 
interferometric facility, which are listed at the end of this section. 
\subsection{Discussion of the general science case for the next 
generation optical/infrared interferometric facility}
Based on the discussion in Sect.~2, it appears essential for
the furthering of our understanding of stellar astrophysics that 
the complete set of fundamental stellar parameters is measured with a 
high precision for all types of stars across the Hertzsprung Russel
diagram. Interferometric techniques have already proven their ability
to derive high-precision fundamental stellar parameters, as outlined
in Sect. 3. These parameters measured so far mostly include the first orders
of intensity profiles or CLVs, i.e. precise radii, the strength of the 
limb-darkening effect, and asymmetric shapes of stellar surfaces. 
These measurements are already augmented with additional 
observational information and compared to theoretical models of 
stellar atmospheres and stellar evolution at a level that challenges 
or confirms theory. However, these measurements are so far mostly limited 
to the apparently largest and brightest stars, and do not provide a 
good sample of the whole Hertzsprung-Russel diagram, i.e. of the 
complete evolutionary tracks. Also, as discussed in Sect. 2, not only 
the basic set of fundamental parameters \{$M$, $L$, $R$\} are of 
fundamental importance for stars in certain evolutionary phases.
Also parameters such as $I$, $B$, $\dot{M}$ may be essential 
(see Sect. 2), and could be characterized (spatially resolved)
with the new facility.

The science case for the next generation optical/infrared 
interferometric facility can thus be summarized as follows:\\[1ex]

``High accuracy measurements of the full set of 
fundamental parameters of stars in all evolutionary 
stages across the Hertzsprung-Russel diagram.''\\[1ex]

Within this general framework, the following topics may be of
particular importance for stellar astrophysics.

\begin{itemize}
\item Observational constraints of the precise effects on the
main sequence.
\item A detailed analysis of metal-poor stars and their evolution.
\item A characterization of low-mass stars.
\item An improved description of high-mass stars, in particular
concerning their mass-loss on the main sequence.
\item A better understanding of the mass-loss process from evolved 
stars, including characteristics of the circumstellar environment and
the formation of asymmetric envelopes (see also Sect. 6).
\item Measurements of magnetically active stars including surface
features.
\item Investigations of the effects of stellar rotation on
stellar evolution and stellar atmospheres.
\item An improved description of convection and turbulent mixing.
\end{itemize}

Observational constraints related to these topics require
high-precision interferometric measurements of stars, including
stars with small angular diameters and low apparent brightness.
They require as well the coupling of interferometric techniques
with high spectral resolution. In order to estimate the
specifications for the interferometric facility needed, we 
consider below some limiting cases. 
\subsection{Limiting cases}
The general science case presented above includes the measurement
of many different stars at all evolutionary phases. The different
aspects discussed imply different specifications for the next-generation
interferometric facility. 

The limiting case with respect to baseline length and collecting area may
be the case of a very low mass star at the bottom of the main sequence.\\
According to recent evolutionary calculations (Chabrier et al. 2000),
a star at the limit to a brown dwarf (mass 0.08\,M$_\odot$) with a typical
age of 5\,Gyr has a radius of 0.1\,R$_\odot$, and a luminosity of
$\log L=$ -3.6\,L$_\odot$. The absolute magnitudes are $M_V \sim 19$,
$M_K \sim 11$, and $M_M \sim 10$. For a close star at a distance of
10\,pc, this corresponds to an angular diameter of 0.1\,mas.
For visibility measurements down to 1\% (without the use of 
bootstrapping), the limiting magnitudes for this star correspond to
$m_V \sim 24$, $m_K \sim 16$, and $M_m \sim 15$. The need to sample the
CLV with at least 3--5 resolution elements (see above) at a
typical wavelength of 1$\mu$m implies a maximum baseline of 6--10\,km.

The shortest baseline needed and the required field of view is maybe
determined by apparently large evolved stars including their 
circumstellar envelope. In order to simultaneously characterize the
stellar surface and the circumstellar environment, scales up to
1000\,R$_\odot$, corresponding at 10\,pc to $\sim$\,1\,arcsec should
ideally be sampled. However, such studies may also be reached by
combination of interferometry with high spatial resolution observations
at single telescopes.

The required spectral resolution is determined by the need to 
characterize magnetic field strengths and abundance ratios
for several resolution elements across the stellar disk. The spectral
resolution should at least reach up to the values currently reached
with single-telescope instruments, i.e. about 100\,000.
\subsection{Specifications and requirements for future interferometers}
Based on the science case presented above, and taking into account
the general aspects discussed in Sect. 2 (in particular the specifications
in Sect. 2.4), the specifications and requirements for 
our next-generation optical/infrared interferometric
facility can be given as follows.

\paragraph{Sensitivity:}
The sensitivity is set by low-mass stars at the bottom of the main
sequence to $m_V \sim 24$, $m_K \sim 16$, and $m_M \sim 15$ for a star
at 10\,pc distance and visibility values down to 1\%.
\paragraph{Observing mode:}
The requirement for high-precision estimates of the
transfer function in order to measure high-precision 
fundamental stellar parameters would benefit from the
use of multi fields, one on the scientific
star, and one on the calibrator.
\paragraph{Spatial resolution:}
The longest baselines are determined by observations with 
3--5 resolution elements
on a 0.1\,mas star, corresponding to a maximum baseline of 6--10\,km
at a typical wavelength of 1$\mu$m. Required are a minimum coverage
of 1--2 radial points per resolution element, i.e. 3--10 $uv$ radii
up to the longest baseline. The azimuth angle should be sufficiently
covered as well by 2--3 samples (plus the use of earth rotation).
This results in 6--30 stations. In order to use them for stars of 
different angular diameter, they should ideally be re-locatable from 
shortest scales of up to 100\,m to largest scales of up to 10\,km.
\paragraph{Wavelength range:}
The comparison to model atmospheres in order to derive high-precision
CLV measurements, and thus accurate radii determinations, implies the
need for a broad wavelength coverage from the optical to the infrared,
i.e. about 0.4\,$\mu$m to 10\,$\mu$m.
\paragraph{Spectral resolution and polarimetric capability:}
The measurement of several of the fundamental parameters
such as surface gravity, metallicity requires a high 
spectral resolution of $>$\,100\,000.
\paragraph{Field of view:}
The field of view would ideally include the circumstellar 
environment. This would imply a value of about 1\,arcsec.
\paragraph{Astrometric precision}
A high astrometric precision is required for 
determinations of astrometric orbits of binary and multiple
systems, which could greatly improve mass estimates (see 
the presentation by F. Verbunt, these proceedings). Also,
the correlation of the astrometric position of the stellar
surface with respect to additional observational information
of the circumstellar environment, as probed for instance
by radio interferometry (see Sect.~6), would be of great value. A precision
of a fraction of the stellar angular diameter would be needed,
i.e. up to the order of 10-100\,$\mu$arcsec.
\section{Synergy effects with external projects}
Interferometric observations with specifications outlined
in Sect. 5 alone will be able to provide important measurements 
of fundamental stellar parameters, and will enlarge our view
of stellar physics. However, the observational determination 
of the complete set of fundamental stellar parameters discussed 
in Sect. 2 will require additional observational information.
As a result, carefully planned synergies with other instrumentation
projects appear valuable. A few options are outlined in the 
following. 
\subsection{Bolometric fluxes}
The primary measurement of effective temperatures of stars
obtaine by combination of measured angular diameters and 
bolometric fluxes. Hence, it would be valuable to combine
the next generation interferometric facility with a facility
that allows us to obtain high-precision bolometric fluxes of 
the same sources, and ideally at the same time. The latter is 
particularly important for variable stars. Also, the measurement 
of high-precision lightcurves of variable stars which are 
targets of the interferometer would be valuable. A carefully
planned synergy with such a facility appears important, so that
the bolometric fluxes of the same sources can easily be
obtained. Currently, measured effective temperatures are often
limited by the measurement of the flux rather than by the measurement
of the angular diameter, in spite of the fact that the former
is in principle easier to obtain.  
\subsection{Distances/parallaxes}
For comparison with theoretical stellar models, absolute
stellar radii are needed rather than angular radii. 
Hence, the knowledge of precise distances is required, in 
particular when higher precision angular diameters are obtained.
Often, these distances are currently not available with the
required precision. Synergies with projects such as 
GAIA or SIM appear interesting.
\subsection{VLBA and ALMA}
Investigations of the mass-loss and the circumstellar environment
of evolved stars benefit from synergies of optical/infrared
interferometry and radio interferometry. These different
techniques at different wavelengths probe different regions of the 
star itself and its circumstellar environment, and thus provide us 
with a more complete picture of the stellar parameters, and in 
particular of the mass-loss process. For instance, 
infrared observations of the stellar photosphere or circumstellar 
dust have already been compared to radio observations of 
circumstellar masers and vice versa 
(e.g., Danchi 1994; Greenhill et al. 1995; 
Cotton et al. 2004; Monnier et al. 2004; Boboltz \& Wittkowski 2005). 
Such approaches are still rare, and furthermore some of these 
studies suffer from combinations of observations of variable stars 
widely spaced in time, thus limiting the accuracy of the comparison.
\begin{figure}
\centering
\resizebox{0.6\hsize}{!}{\includegraphics{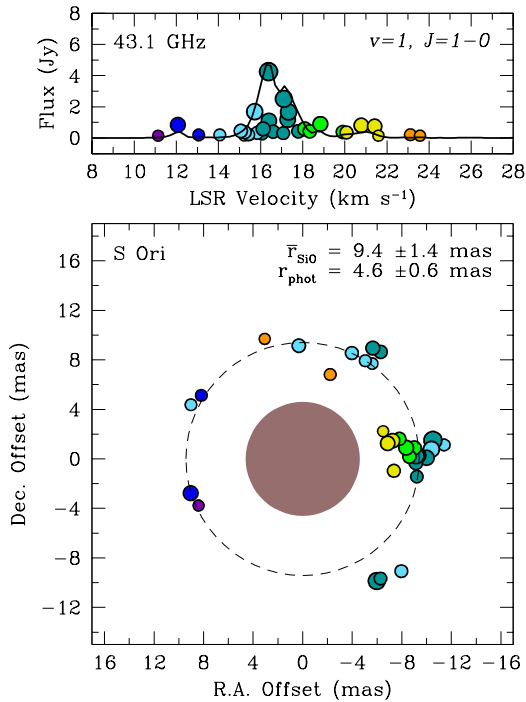}}
\caption{VLBA observation of the 43.1 GHz maser emission toward 
S Ori (color-coded in bins of radial velocity) concurrent with VLTI 
observations of the stellar photosphere at near-infrared wavelengths 
(represented by the disk in the center of the distribution). From
Boboltz \& Wittkowski (2005).}
\label{fig:sori}
\end{figure}
Boboltz \& Wittkowski (2005) conducted concurrent observations of 
the Mira variable S Ori as part of a program designed to use the power 
of long baseline interferometry at infrared and radio wavelengths to 
study the photospheres and nearby circumstellar envelopes of 
evolved stars. Figure~\ref{fig:sori} shows the results of the 
first ever coordinated observations between 
NRAO's VLBA (Very Long Baseline Array) and 
ESO's VLTI (Very Large Telescope Interferometer) facilities. 
The VLBA was used to observe the 43-GHz SiO maser emission 
(represented by the circles color-coded in bins of radial velocity) 
concurrent with VLTI observations of the stellar photosphere at 
near-infrared wavelengths (represented by the red disk in the center 
of the distribution).  The SiO masers were found to lie at a distance 
of roughly 1.7 stellar radii or 1.5 AU.
With concurrent observations such as these can parameters of 
the circumstellar gas, as traced by the SiO masers, be related to 
the star itself at a particular phase in its pulsation cycle. Such 
measurements can be used to constrain models of stellar pulsation, 
envelope chemistry, maser generation, and stellar evolution.

Synergies of a next-generation optical/infrared interferometer
with VLBA appear interesting as well. Current VLTI observations
can not match the maximum angular resolution of radio long-baseline
interferometry.
\begin{figure}
\centering
\resizebox{0.7\hsize}{!}{\includegraphics{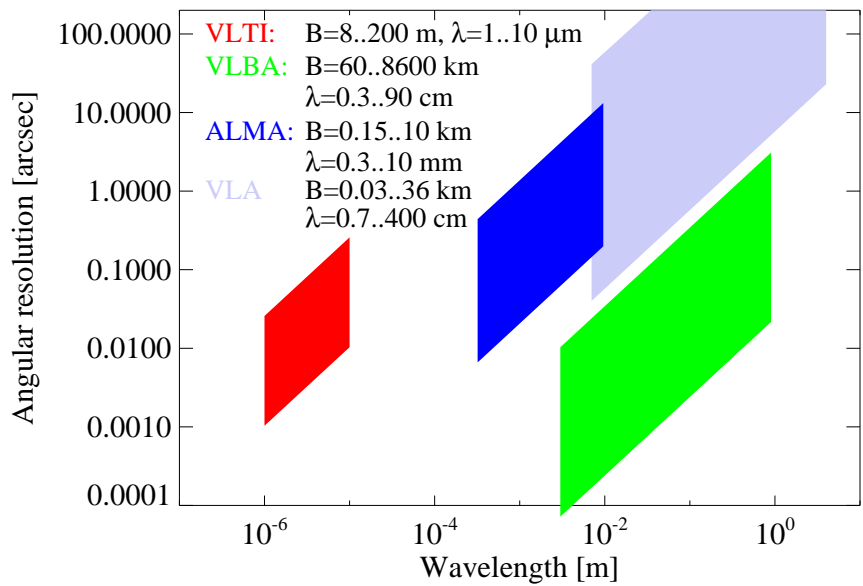}}
\caption{Comparison of resolution and wavelength ranges
of the infrared, millimeter, and radio interferometric facilities 
VLTI, ALMA, VLBA, and VLA. Radio interferometry with the VLBA reaches
currently an angular resolution which is about a factor of 10 above 
that of the VLTI. Transcontinental radio VLBI reaches even 
better angular resolutions. ALMA is restricted to lower angular 
resolution, but also typically probes cool dust, that is usually 
distributed at larger scales. Coordinated interferometric observations 
at infrared and radio wavelengths, using for instance the VLTI and the
VLBA (Boboltz \& Wittkowski 2005) in the field of stellar surfaces
and circumstellar environment, have already started. More detailed
multi-wavelength studies, also including mm interferometry with ALMA,
are upcoming. A next-generation optical interferometric facility with 
baselines of 1-10\,km would clearly enhance the possibilities of such
approaches in stellar astrophysics. The maximum angular resolution 
would be comparable to that of current radio interferometry, and also 
the image fidelity could better match that of the 10-station VLBA or 
even the 64-station ALMA.}
\label{fig:comp}
\end{figure}
Fig.~\ref{fig:comp} shows a comparison of VLTI, ALMA, VLBA, and VLA
in terms of wavelengths and angular resolution. Radio interferometry 
with the VLBA reaches currently an angular resolution which is 
about a factor of 10 above that of the VLTI. Transcontinental radio 
VLBI reaches even better angular resolutions. A next-generation
optical/infrared interferometer could reach the maximum
angular resolution currently provided by the VLBA, and thus allow us
to compare VLBA and infrared interferometric results in more detail.
Also, the image fidelity that is currently provided by the 10-station
VLBA will not easily be reached by VLTI, but can be reached with
the next generation optical/infrared interferometer.

Similar synergies with ALMA, the upcoming millimeter interferometer,
are interesting as well. ALMA is restricted to lower angular 
resolution, but also typically probes cool dust, that is usually 
distributed at larger scales.

Synergies of infrared and radio interferometry may not only
be interesting for evolved stars as outlined above, but also for
other aspects of stellar physics, such as star formation.
\subsection{Coordination with theoretical efforts}
Finally, it seems appealing to coordinate observational and theoretical
efforts in the field of stellar astrophysics in a more coordinated
and closer way. With increasing observational capabilities and improved
measurements of fundamental stellar parameters, also their interpretation
and their comparison to model prediction will need to become more 
extensive and more sophisticated. Furthermore, the planning of 
observations will benefit more and more from simulations using
theoretical models. Model studies to investigate the
observational precisions needed and identifying the most interesting
measurements in terms of, for instance, targets and wavelength
ranges, can ideally be conducted already in parallel to the planning
of a next-generation observational facility. 
%
%
\section*{Acknowledgements}
I am grateful for valuable discussions with T. Driebe and M. Petr-Gotzens.
%
%
\beginrefer
\refer Andersen, J.\ 1991, A\&ARv, 3, 91 

\refer Antoniucci, S, Paresce, F., \& Wittkowski, M. 2004, A\&A, in press

\refer di Benedetto, G. P., \& Foy, R. 1986, A\&A, 166, 204

\refer Boboltz, D., \& Wittkowski, M. 2005, ApJ, in press 

\refer Burns, D., Baldwin, J. E., Boysen, R. C., et al. 1997, MNRAS, 290, L11

\refer Chabrier, G., Baraffe, I., Allard, F., \& Hauschildt, P.\ 2000, 
ApJ, 542, 464

\refer Cotton, W.~D., et al.\ 2004, A\&A, 414, 275

\refer Danchi, W. C., et al. 1994, AJ, 107, 1469

\refer Domiciano de Souza, A., Kervella, P., Jankov, S., Abe, L., 
Vakili, F., di Folco, E., \& Paresce, F.\ 2003, A\&A, 407, L47 

\refer Fedele, D., Wittkowski, M., Paresce, F., et al. 2005, A\&A, in press 

\refer Girardi, L., Bressan, A., Bertelli, G., \& Chiosi, C.\ 2000, 
A\&AS, 141, 371 

\refer Greenhill, L. J. et al. 1995, ApJ, 449, 365

\refer Hanbury Brown, R., Davis, J., Lake, R. J. W.,
Thompson, R.J. 1974, MNRAS, 167, 475

\refer Haniff, C. A., Scholz, M., \& Tuthill, P. G. 1995, MNRAS, 276, 640

\refer Hofmann, K.-H., Scholz, M., \&  Wood, P.R.\ 1998, A\&A, 339, 846

\refer Ireland, M. J., Scholz, M., Wood, P. R.\ 2004a, MNRAS, 352, 318

\refer Ireland, M.~J., Scholz, M., Tuthill, P.~G., \& Wood. P.~R.\ 2004b, 
MNRAS, in press

\refer Jeffries, R.~D.\ 1997, MNRAS, 288, 585

\refer Maeder, A.~\& Meynet, G.\ 2003, A\&A, 411, 543

\refer Maeder, A.~\& Meynet, G.\ 2004, A\&A, 422, 225 

\refer Monnier, J.~D.\ 2003, Reports of Progress in Physics, 66, 789 

\refer Monnier, J. D., Millan-Gabet, R., Tuthill, P. G., et al. 2004, 
ApJ, 605, 436 

\refer Ohnaka, K., Bergeat, J, Driebe, T., et al. 2005, A\&A, in press

\refer Pasquini, L., Bonifacio, P., Randich, S., Galli, D., \& 
Gratton, R.~G.\ 2004, A\&A, 426, 651 

\refer Perrin G., Coud\'e du Foresto, V., Ridgway, S. T., et al. 1999, A\&A,
345, 221

\refer Perrin, G. et al. 2004, A\&A, 426, 279 

\refer Quirrenbach, A., Mozurkewich, D., Buscher, D. F.,
Hummel, C. A., Armstrong, J.T. 1996, A\&A, 312, 160

\refer Quirrenbach, A.\ 2001, ARA\&A, 39, 353 

\refer Scholz, M.\ 1998, IAU Symp.~189: Fundamental Stellar Properties, 
189, 51

\refer Scholz, M.\ 2001, MNRAS, 321, 347

\refer Scholz, M.\ 2003, Proc. SPIE, 4838, 163

\refer Scholz, M., \& Wood, P. R., 2004, private communication

\refer S{\' e}gransan, D., Kervella, P., Forveille, T., \& 
Queloz, D.\ 2003, A\&A, 397, L5 

\refer Tej, A., Lan{\c c}on, A., Scholz, M., \& Wood, P.~R.\ 2003b, A\&A, 
412, 481

\refer van Belle, G.~T., Ciardi, D.~R., Thompson, R.~R., Akeson, R.~L., 
\& Lada, E.~A.\ 2001, ApJ, 559, 1155 

\refer Wittkowski, M., Langer, N., \& Weigelt, G. 1998, A\&A, 340, L39

\refer Wittkowski, M., Hummel, C. A., Johnston, K. J., et al. 2001, 
A\&A, 377, 981

\refer Wittkowski, M., Aufdenberg, J., \& Kervella, P. 2004, A\&A, 413, 711

\refer Woodruff, H. C., Eberhardt, M., Driebe, T., et al. 2004, A\&A, 
421, 703

\refer Yi, S.~K., Kim, Y., \& Demarque, P.\ 2003, ApJS, 144, 259

\endrefer           
\end{document}